\documentstyle[12pt]{article}
\begin{document}

\begin{center}
Probability representation entropy for spin-state tomogram.\\
O.~V.~Man'ko, and V.~I.~Man'ko\\
Lebedev Physical Institute, Moscow, Leninskii pr., 53
\end{center}

\begin{abstract}
Probability representation entropy (tomographic entropy) 
of arbitrary quantum state is introduced. 
Using the properties of spin tomogram to be standard
probability distribution function the tomographic entropy
notion is discussed. Relation of the tomographic entropy to
Shannon entropy and von Neumann entropy is elucidated.
\end{abstract}

\section{Introduction}
There exists formulation of quantum mechanics 
(see, e.g.~\cite{Mancini1,36}) where quantum 
system states are associated with probabilities (instead of 
wave functions or density matrix). One can use this formulation 
to develop naturally notion of entropies related to the 
probabilities. The probabilities determining quantum states 
are called state tomograms. Since arbitrary quantum state of any 
quantum system is determined by the probability the entropy (which 
we call probability representation entropy) is always associated 
with the quantum state. In~\cite{Dod,JETP} tomograms of spin 
states were introduced. The star-product formalism for tomographic 
symbols of spin operators was introduced 
in~\cite{MarmoPhysicaScripta,MarmoJPhys}. Spin tomography was
developped in~\cite{36,ManciniTombesi,KlimovSmirnov,AndreevOlga,Safonov}.
The tomography for two spins was introduced
in~\cite{SafonovVI,AndreevVIJETP}. The tomograms for spin were also
considered in~\cite{Weigert,castanos,Leonard}. Since tomograms
of quantum states are standard probability distribution functions,
all the characteristics of the probability distributions known in
probability theory can be applied also in the framework of tomographic
probability representation of quantum states. The most important
characteristics related to probability distributions are entropy and
information  (see, e.g.~\cite{Wentzel}). For symplectic
tomograms~\cite{ManciniJOPT95} the notion of entropy has been shortly discussed
in~\cite{MAMankoJRLR}.

The aim of this work is to introduce the notion of tomographic entropy and
information and to consider these notion for spin states. We consider both one particle and multiparticle
cases. In classical probability theory the notion of Shannon
entropy~\cite{Shannon} is the basic one. In quantum mechanics von
Neumann~\cite{vonNeumann} introduced entropy related to density operator
(see, e.g.~\cite{Holevo}). We will obtain the relation of the introduced 
tomographic entropy and information to the Shannon and von Neumann entropies. 
The article is organised as follows.
In Section II the standard notions of probability theory are reviewed. 
In Section III the spin tomography of one particle is reviewed and entropy 
of spin states is discussed. In Section IV spin tomograms for two particles 
are discussed and corresponding entropies are introduced. In Section V the 
relations of tomographic entropies to von Neumann entropy are discussed. 
Conclusions and perspectives are presented in Section VI.

\section{Properties of entropy}

In this section we review standard properties of entropy following presentation and 
notation of probability theory given in~\cite{Wentzel}. In quantum mechanics a pure 
state is associated with a vector $|\psi\rangle$ in the Hilbert space. A mixed state 
is associated with a density matrix $\rho$. In probability representation of quantum 
mechanics one describes the state by a probability distribution depending on extra 
parameters. This probability distribution is either function of discrete random 
variable (spin) or a function of continuous random variable (position). Extra 
parameters determine reference frame where these variables are measured. 
Thus one has one-to-one correspondence (invertable map) 
\begin{equation}\label{eq.a}
\rho\leftrightarrow\omega
\end{equation}
where $\omega$ is the discussed probability distribution. With each probability 
distribution one can associate entropy following the Shannon prescription. For 
any quantum state (given by $\omega$) we associate entropy $S$. This entropy 
$S$ we call probability representation entropy of quantum state. This entropy is 
the function of discussed extra parameters (reference frame parameters). Below we review 
generic properties of an entropy $H$ used in probability theory. 

Given discrete random variable of the system $X$, which has states $x_i$. The 
Shannon entropy $H(X)$ is defined as 
\begin{equation}\label{eq.C1} 
H(X)=-\sum_{i=1}^n P_i \ln P_i, 
\end{equation}
where $P_i$ is probability distribution function for the random variable (probability 
of the state $x_i$). Entropy of a bipartite system with discrete random variables describing 
subsystems $X$ and $Y$ is defined as 
\begin{equation}\label{eq.C2}
H(X,Y)=-\sum_{i=1}^n\sum_{j=1}^m P_{ij}\ln P_{ij}, 
\end{equation}
where $P_{ij}$ is joint probability distribution of two random variables (joint probabilities 
to have states $x_i$ and $y_j$). Conditional entropy is defined as 
\begin{equation}\label{eq.C3}
H(Y|x_i)=-\sum_{j=1}^m P(y_j|x_i)\ln P(y_j|x_i). 
\end{equation}
Here $P(y_j|x_i)$ is probability for system $Y$ to be in state $y_j$, if the system $X$ is in 
the state $x_i$. Let us define the complete conditional entropy by relation 
\begin{equation}\label{eq.C3a}
H(Y|X)=-\sum_{i-1}^n \sum_{j=1}^m P_{i j}\ln P(y_j|x_i). 
\end{equation}
It is known that 
\begin{equation}\label{eq.C4}
H(X,Y)=H(X)+H(Y|X),\quad H(X,Y)\leq H(X). 
\end{equation}
Also 
\begin{equation}\label{eq.C5}
H(X,Y)\leq H(X)+H(Y) 
\end{equation}
Information on the system $X$ obtained due to observation of the system $Y$ 
is defined by the relation 
\begin{equation}\label{eq.C6}
I_{Y\rightarrow X }=H(X)-H(X|Y).
\end{equation}
One has definition of complete mutual information obtained on systems $X$ and $Y$ 
\begin{equation}\label{eq.C7}
I_{X\leftrightarrow Y}=I_{X\rightarrow Y}=I_{Y\rightarrow X}. 
\end{equation}
Also the information is given by the formula
\begin{equation}\label{eq.C8}
I_{Y\leftrightarrow X}=H(X)+H(Y)-H(X,Y)\geq 0
\end{equation}
Let us apply these general relations to the spin systems using tomographic 
probabilities. For bipartite spin system the subsystem $X$ is identified with 
first particle with spin $j_1$ and the subsystem $Y$ with second particle with spin 
$j_2$ for given group element U(n) which determines the basis in the space 
of spin quantum states. The measured variables $x_i,\,y_j$ are spin projections on 
$z$-axis $m_1$ and $m_2$, respectively. For spin system we denote entropy by 
capital letter S.

\section{Spin states in the tomographic representation} 

According to formalism of probability representation of quantum spin 
states the state of system 
with spin $j$ with density matrix $\rho$ is associated with probability 
distribution function $\omega^{(j)}(m,\vec{n})$, where $m=-j,-j+1,...,j$, the unit 
vector $\vec{n}=(\sin\theta\cos\phi,\sin\theta\sin\phi,\cos\theta)$ determines 
the point on the sphere. This probability distribution (called spin state 
tomogram) was introduced in~\cite{Dod,JETP} and it is normalized for each vector $\vec{n}$, i.e.
\begin{equation}\label{eq.1}
\sum_{m=-j}^{j}\omega(m,\vec{n})=1. 
\end{equation}
It is related to density operator $\rho$ by the formula 
\begin{equation}\label{eq.2}
\omega(m,\vec{n})=\langle j m |D^{(j)+}(u)\rho D^{(j)}(u)|j m \rangle.
\end{equation}
Here states $|j m\rangle$ are standard states with spin projection $m$ 
on the $z$-axis, i.e., 
\begin{equation}\label{eq.3}
\hat j_z|j m\rangle = m|j m\rangle. 
\end{equation}
The matrix $D^{(j)}(u)$ is the matrix of irreducible representation 
of the SU(2) group. The 2x2 - matrix $u$ is the element of this group. We 
use Euler angles $\phi$, $\theta$, $\psi$ as parameters of this element $u$. 
\begin{equation}\label{eq.4}
u(\phi,\theta, \psi)=\pmatrix{
\cos({\theta\over2})e^{\frac{i}{2}(\phi+\psi)}&
\sin({\theta\over2})e^{-{i\over2}(\phi-\psi)}\cr 
-\sin({\theta\over2})e^{-{i\over2}(\phi-\psi)}
&\cos({\theta\over2})e^{-\frac{i}{2}(\phi+\psi)}}. 
\end{equation}
Due to the structure of the formula~(\ref{eq.2}) the tomogram does not depend 
on Euler angle $\psi$. The density matrix $\rho^{(j)}$ with matrix elements 
$\rho^{(j)}_{mm'}$ can be constructed if one knows the tomogram according 
to relation~\cite{JETP} 
\begin{eqnarray}\label{eq.5}
&(-1)^{m'}\,\rho_{m m'}^{(j)}\,=\,&\sum_{k=0}^{2j}\,\sum_{l=-k}^{k}\,
(2k+1)^2\sum_{i=-j}^{j} 
\int (-1)^{i}\,\omega( i,u)\,D_{0\,l}^{k}
\left(u\right)\nonumber\\
&\times&\pmatrix{j&j&k\cr
i&-i&0}\,\pmatrix{j&j&k\cr m&-m'&l}\,
\frac{d\Omega}{8\,\pi^2}\,,  
\end{eqnarray}
Here $m,\,m'=-j,-j+1,...,j$; the integration is performed with 
respect to the angular variables $\phi\,,\theta$ and $\psi$, i.e.,
$$\int d\Omega=\int_0^{2\pi}d\phi\int_0^{2\pi}d\psi\int_0^{\pi}
\sin\theta d\theta.$$
The Shannon entropy~\cite{Shannon} associated to any probability 
distribution function provides the entropy associated to spin tomogram. 
Thus we define the tomographic entropy as the function on the sphere 
\begin{equation}\label{eq.6} 
S(\vec{n})=-\sum_{m=-j}^{j}\omega(m,\vec{n})\ln\omega(m,\vec{n}).
\end{equation}
The tomography of quark states suggested in~\cite{KlimovSmirnov} 
was generalized and the unitary spin-tomography 
was discussed in~\cite{Sudarshan}. The unitary spin tomogram is defined 
as follows 
\begin{equation}\label{eq.7}
\omega\left(m,U(n)\right)=\langle j m|U^+(n)\rho 
U(n)|j m\rangle.
\end{equation}
Here $n=2j+1$, $n$-dimensional matrix $U(n)$ is element of unitary group. 
Since $D^{(j)}(u)$ is subgroup of the unitary group, the tomogram~(\ref{eq.7}) 
determines the density matrix of spin state. If one takes as matrix $U(n)$ 
the matrix $D^{(j)}(u)$ the unitary spin tomogram becomes the spin tomogram. The 
function~(\ref{eq.7}) is the probability distribution function which is 
normalised for each group element $U(n)$, i.e.,
\begin{equation}\label{eq.8}
\sum_{m=-j}^{j}\omega\left(m,U(n)\right)=1. 
\end{equation}
In fact the tomogram is probability to get the spin projection $m$ in the 
"rotated" basis 
\begin{equation}\label{eq.9}
|j \tilde m\rangle=U(n)|j m\rangle. 
\end{equation}
The application of Shannon entropy formalism to the unitary spin tomogram 
defines the unitary tomographic entropy which is the function on unitary 
group, i.e., 
\begin{equation}\label{eq.10}
S\left(U(n)\right)=-\sum_{m=-j}^j\omega\left(m,U(n)\right)
\ln\omega\left(m,U(n)\right). 
\end{equation}
For $U(n)$ taken as $D^{(j)}(u)$ the entropy~(\ref{eq.10}) coincides with 
entropy~(\ref{eq.6}). 

\section{Tomography for two-spin particle}

Below we apply the results discussed in Section 2 for two subsystems $X$ 
and $Y$ to the case of two spins. Let us consider now two particles with 
spin $j_1$ and $j_2$, respectively. The basis in the space of states can 
be given by product vector 
\begin{equation}\label{eq.11}
|m_1 m_2\rangle=|j_1 m_1\rangle|j_2 m_2\rangle.
\end{equation}
The density matrix $\rho$ of a system state can be mapped onto 
spin-tomogram~\cite{Safonov,Sudarshan}
\begin{equation}\label{eq.12}
\omega(m_1,m_2,\vec{n_1},\vec{n_2}) =\langle m_1 m_2|D^{(j_1)+}(u_1)
\otimes D^{j_2)+}(u_2)\rho D^{(j_2)}(u_2)\otimes 
D^{(j_1)}(u_1)|m_1 m_2\rangle. 
\end{equation} 
This is joint probability distribution function for two discrete 
random variables $m_1$ and $m_2$ which are spin projections on the 
directions $\vec{n_1}$ and $\vec{n_2}$, respectively. The function 
is normalised 
\begin{equation}\label{eq.13}
\sum_{m_1=-j_1}^{j_1}\sum_{m_2=-j_2}^{j_2}
\omega(m_1,m_2,\vec{n_1},\vec{n_2})=1.
\end{equation} 
Tomographic entropy  $S(\vec{n_1},\vec{n_2})$ can be associated with 
this probability distribution function 
\begin{equation}\label{eq.14}
S(\vec{n_1},\vec{n_2})=-\sum_{m_1=-j_1}^{j_1}\sum_{m_2=-j_2}^{j_2}
\omega(m_1,m_2,\vec{n_1},\vec{n_2})
\ln\omega(m_1,m_2,\vec{n_1},\vec{n_2}). 
\end{equation}
This tomographic entropy depends on the points on two spheres 
determined by unit vectors $\vec{n_1}$ and $\vec{n_2}$. The 
tomographic probability for two particles determines the 
tomographic probability for one particle, e.g., 
\begin{equation}\label{eq.15}
\omega(m_1,\vec{n_1})=\sum_{m_2=-j_2}^{j_2}
\omega(m_1,m_2,\vec{n_1},\vec{n_2}). 
\end{equation}
In~\cite{Sudarshan} unitary spin tomogram was introduced by relation 
\begin{equation}\label{eq.16}
\omega(m_1,m_2,U(n))=\langle m_1 m_2|U^+(n)\rho U(n)|m_1 m_2\rangle,\,
n=(2j_1+1)(2j_2+1). 
\end{equation}
This tomogram is joint probability distribution function depending 
on unitary group element $U(n)$. It is normalised for each group 
element, i.e., 
\begin{equation}\label{eq.17}
\sum_{m_1=j_1}^{j_1}\sum_{m_2=-j_2}^{j_2}\omega\left(m_1, m_2,U(n)\right)=1. 
\end{equation}
The joint tomographic probability~(\ref{eq.16}) determines the 
tomographic probability for one particle depending on unitary group 
element 
\begin{equation}\label{eq.18}
\omega_1\left(m_1, U(n)\right)=\sum_{m_2=-j_2}^{j_2}
\omega\left(m_1,m_2,U(n)\right). 
\end{equation}
Analogous probability can be obtained for the second spin. We can 
associate with the tomographic probability the entropy which 
is the function on the unitary group  
\begin{equation}\label{eq.19}
S_1(U(n))=-\sum_{m_1=-j_1}^{j_1}\omega\left(m_1,U(n)\right)
\ln\omega\left(m_1,U(n)\right).
\end{equation}
Also the tomographic entropy related to the tomogram~(\ref{eq.16})
determines  the tomographic probability for one particle depending 
on unitary group element 
\begin{equation}\label{eq.20}
S(U(n))=-\sum_{m_1=-j_1}^{j_1}\sum_{m_2=-j_2}^{j_2}
\omega\left(m_1,m_2,U(n)\right)
\ln\omega\left(m_1,m_2,U(n)\right).
\end{equation}
This entropy depends on unitaty group parameters. For the matrix 
$U(n)=D^{(j_1)}(u_1)\otimes D^{(j_2)}(u_2)$ the entropy~(\ref{eq.20}) 
coincides with the entropy~(\ref{eq.14}). 

We can construct also conditional probability distribution for first spin 
\begin{equation}\label{eq.20a}
\omega_{C1}\left(m_1,U(n)\right)=\frac{\omega\left(m_1,m_2, U(n)\right)}
{\sum_{m_1=-j_1}^{j_1}\omega\left(m_1,m_2,U(n)\right)}. 
\end{equation}
Analogous formula can be written for second spin. Tomographic information 
$I_{Y\leftrightarrow X}$, where $X$ is $j_1$- spin system and $Y$ is 
$j_2$ - spin system is given by the function on the unitary group 
\begin{equation}\label{eq.20b}
I_{j_1\leftrightarrow j_2}\left(U(n)\right)=\sum_{m_1=-j_1}^{j_1}
\sum_{m_2=-j_2}^{j_2}\omega\left(m_1,m_2,U(n)\right) 
\ln{\frac{\omega\left(m_1,m_2, U(n)\right)}
{\omega_1\left(m_1,U(n)\right)\omega_2\left(m_2,U(n)\right)}}.
\end{equation}
Here $\omega_2\left(m_2, U(n)\right)$ is given by~(\ref{eq.18}) with 
replacement $1\leftrightarrow 2$.

\section{Information and Relation to von Neumann entropy}

The unitary spin entropy~(\ref{eq.10}) for the case of spin state 
of single particle defines the von Neumann entropy of this state 
\begin{equation}\label{eq.21} 
S_N=-Tr\left[\rho\ln\rho\right].
\end{equation}
In fact, there exist elements of unitary group $U^{(0)}(n)$, $(n=2j+1)$, 
which diagonalize the density matrix $\rho$. For these elements $U^{(0)}(n)$ 
the tomogram is equal to probability distribution function which 
coincides exactly with eigenvalues of the density matrix. It means that 
the tomographic  entropy~(\ref{eq.10}) for these values of unitary group 
elements is equal to von Neumann entropy of the spin state, i.e.,
\begin{equation}\label{eq.22}
S(U^{(0)}(n))=S_N.
\end{equation}
On the other hand it is obvious that for the elements $U^{(0)}(n)$ the 
tomographic entropy takes minimal possible value. It follows from the 
property that the probability distributions determined by density matrix 
diagonal elements for the matrix obtained by means of unitary rotation 
of the basis are smoother than distributions provided by eigenvalues of the 
density matrix~\cite{Sudarshan}. Thus the von Neumann entropy is the minimum 
of tomographic entropy. 

One can introduce complete mutual information for two spins using the 
relation 
\begin{equation}\label{eq.23} 
I_{j_2\leftrightarrow j_1}\left(U(n)\right)=S_1\left(U(n)\right)+
S_2\left(U(n)\right)-S\left(U(n)\right).
\end{equation}
Here the tomographic entropies are functions on the unitary group and the 
introduced information is also the function on the unitary group. The 
discussed entropies and information can be used to study the properties of 
multipartite quantum states from new viewpoint. 

\section{Conclusion}
We summarize the main results of the work. We introduced concept of 
probability representation entropy of quantum states. This entropy (we 
also call this entropy tomographic entropy) is defined via tomographic 
probability distribution function determining the quantum state in the 
approach which is called probability representation of quantum mechanics. 
The tomographic probability was defined using standard Shannon relation 
of entropy and probability distribution. We applied this approach to spin 
systems. Since for spin system the tomographic probability of quantum state 
is the function on unitary group the tomographic entropy is also the 
function on unitary group. The von Neumann entropy of quantum state was shown 
to be the minimal value of the tomographic entropy. This minimum of 
tomographic entropy is realised for a set of unitary group elements which are 
diagonal unitary matrices commuting with density matrix considered in the 
basis where it is also diagonal. Using standard relation of entropy and 
information the notion of probability representation information (tomographic 
information) was introduced. Some standard relations known in probability 
theory were applied to the tomographic entropy and tomographic information 
of quantum states. The construction of tomographic entropies and informations 
for multipartite system is straightforward. 
The introduced notions of tomographic entropy and tomographic 
information elucidate new aspects of quantum states which are naturally appear 
in probability represenation of quantum mechanics*.
  
----------------------------------------------------------------------------

*When the paper was finished authors became aware of~\cite{Pom,Popescu}, 
where some aspects of other entropies related to discussed ones in this  
work were considered (without the tomographic framework). 

-----------------------------------------------------------------------------

\section{Acknowledgement}
This study was supported by the Russian Foundation for Basic Research under 
Project No 03-02-16408.

\end{document}